\documentclass[twocolumn]{revtex4}
\usepackage{amsmath,amssymb}
\usepackage{graphicx}
\usepackage{dcolumn}
\usepackage[vcentermath]{youngtab}
\usepackage[dvips]{epsfig,color}
\usepackage{bm}

\begin{document}
\title{The unquenched quark model\footnote{Presented at EEF70, Workshop on Unquenched Hadron Spectroscopy: Non-Perturbative Models and Methods of QCD vs. Experiment on the occasion of Eef van Beveren's 70th birthday, 1-5 September 2014, University of Coimbra, Portugal}}
\author{E. Santopinto\footnote{email: santopinto@ge.infn.it}}
\affiliation{Dipartimento di Fisica and INFN, Universit\`a di Genova, via Dodecaneso 33, I-16146 Genova, Italy}
\author{R. Bijker, H. Garc{\'{\i}}a-Tecocoatzi}
\affiliation{Instituto de Ciencias Nucleares, Universidad Nacional Aut\'onoma de M\'exico, AP 70-543, 04510 M\'exico DF, M\'exico}
\author{J. Ferretti}
\affiliation{Dipartimento di Fisica and INFN, Universit\`a di Roma "Sapienza", Piazzale A. Moro 5, I-00185 Roma, Italy}

\begin{abstract}
In this contribution, we briefly analyze the formalism of the unquenched quark model (UQM) and its application to the calculation of hadron spectra with self-energy corrections, due to the coupling to the meson-meson continuum. In the UQM, the effects of $q \bar q$ sea pairs are introduced explicitly into the quark model through a QCD-inspired $^{3}P_0$ pair-creation mechanism. The UQM formalism can be extended to include also the effects of hybrid mesons, i.e. hybrid loops. The main applications to spectroscopy and decays are analyzed. 
\end{abstract}

\maketitle
  
\section{Introduction}
The quark model can reproduce the behavior of observables such as the spectrum and the magnetic moments, but it neglects pair-creation (or continuum-coupling) effects.
Above threshold, this coupling leads to strong decays; below threshold, it leads to virtual $q \bar q - q \bar q$ ($qqq - q \bar q$) components in the hadron wave function and shifts of the physical mass with respect to the bare mass. 
The unquenching of the quark model for hadrons is a way to take these components into account.

Pioneering work on the unquenching of meson quark models was done by Van Beveren and Rupp used an t-matrix approach \cite{vanBeveren:1979bd,vanBeveren:1986ea} , while T\"ornqvist and collaborators \cite{Ono:1983rd,Tornqvist} used their unitarized QM.
These methods were used (with a few variations) by several authors to study the influence of the meson-meson (meson-baryon) continuum on meson (baryon) observables. 
As an example we mention the study of the scalar meson nonet ($a_0$, $f_0$, etc.) of Ref. \cite{vanBeveren:1986ea,Tornqvist:1995kr} in which the loop contributions are given by the hadronic intermediate states that each meson can access. It is via these hadronic loops that the bare states become ``dressed'' and  the hadronic loop contributions totally dominante the dynamics of the process. A very similar approach was developed by Boglione and Pennington in Ref. \cite{Pennington:2002}, in which they investigated the dynamical generation of the scalar mesons by initially inserting only one ``bare seed''. The study of Ref. \cite{Geiger:1989yc} demonstrates that the effects of the $q \bar q$ sea pairs in meson spectroscopy is simply a renormalization of the meson string tension. The strangeness content of the nucleon electromagnetic form factors was investigated in  \cite{Geiger:1996re,Bijker:2012zza}. Also Capstick and Morel in Ref. \cite{Capstick} analyzed  
baryon meson loop effects on the spectrum of nonstrange baryons.   
Eichten {\it et al.} explored the influence of the open-charm channels on the charmonium properties, using the Cornell coupled-channel model \cite{Eichten:1974af} to assess departures from the single-channel  potential-model expectations.
The flavor asymmetry of the proton was studied in the framework of the unquenched quark model (UQM)  \cite{Bijker:2012zza,Bijker:2009up,bottomonium,charmonium,Ferretti:2013vua,Ferretti:2014xqa}, in wich the effects of the quark-antiquark pairs were introduced  into the constituent quark model (CQM) in a systematic way and the wave fuctions were given explicitly. The approach is a generalization of the unitarized quark model \cite{vanBeveren:1979bd,vanBeveren:1986ea,Tornqvist,Tornqvist:1995kr}. 
  
In this contribution, we discuss some of the latest applications of the UQM to the study of meson observables.	

\section{UQM }
\subsection{Formalism} 
\label{Sec:formalism}
In the unquenched quark model for baryons \cite{Bijker:2012zza,Bijker:2009up} and mesons \cite{bottomonium,charmonium,Ferretti:2013vua,Ferretti:2014xqa}, the hadron wave function is made up of a zeroth order $qqq$ ($q \bar q$) configuration plus a sum over the possible higher Fock components, due to the creation of $^{3}P_0$ $q \bar q$ pairs. Thus,  we have 
\begin{eqnarray} 
	\label{eqn:Psi-A}
	\mid \psi_A \rangle &=& {\cal N} \left[ \mid A \rangle 
	+ \sum_{BC \ell J} \int d \vec{K} \, k^2 dk \, \mid BC \ell J;\vec{K} k \rangle \right.
	\nonumber\\
	&& \hspace{2cm} \left.  \frac{ \langle BC \ell J;\vec{K} k \mid T^{\dagger} \mid A \rangle } 
	{E_a - E_b - E_c} \right] ~, 
\end{eqnarray}
where $T^{\dagger}$ stands for the $^{3}P_0$ quark-antiquark pair-creation operator \cite{bottomonium,charmonium,Ferretti:2013vua,Ferretti:2014xqa}, $A$ is the baryon/meson, $B$ and $C$ represent the intermediate state hadrons, $E_a$, $E_b$ and $E_c$ are the corresponding energies, $k$ and $\ell$ the relative radial momentum and orbital angular momentum between $B$ and $C$ and $\vec{J} = \vec{J}_b + \vec{J}_c + \vec{\ell}$ is the total angular momentum. 
It is worthwhile noting that in Refs. \cite{bottomonium,charmonium,Ferretti:2013vua,Ferretti:2014xqa,Kalashnikova:2005ui}, the constant pair-creation strength in the operator (\ref{eqn:Psi-A}) was substituted with an effective one, to suppress unphysical heavy quark pair-creation. 

In the UQM \cite{Bijker:2012zza,Bijker:2009up,bottomonium,charmonium,Ferretti:2013vua,Ferretti:2014xqa}, the matrix elements of an observable $\hat O$ can be calculated as
\begin{equation}
	O = \left\langle \psi_A \right| \hat O \left| \psi_A \right\rangle \mbox{ }, 
\end{equation}
where $\left| \psi_A \right\rangle$ is the state of Eq. (\ref{eqn:Psi-A}). 
The result will receive a contribution from the valence part and one from the continuum component, which is absent in naive QM calculations. 
 
The introduction of continuum effects in the QM can thus be essential to study observables that only depend on $q \bar q$ sea pairs, like the strangeness content of the nucleon electromagnetic form factors \cite{Geiger:1996re,Bijker:2012zza}. 
In other cases, continuum effects can provide important corrections to baryon/meson observables, like the self-energy corrections to meson masses \cite{bottomonium,charmonium,Ferretti:2013vua,Ferretti:2014xqa} or the importance of the orbital angular momentum in the spin of the proton \cite{Bijker:2009up}. 

\subsection{$c \bar c$ and $b \bar b$ spectra with self-energy corrections in the UQM}
In Refs. \cite{bottomonium,charmonium,Ferretti:2013vua,Ferretti:2014xqa}, the method was used by some of us to compute the $c \bar c$ and $b \bar b$ spectra with self-energy corrections, due to continuum coupling effects. 
In the UQM, the physical mass of a meson, 
\begin{equation}
	\label{eqn:self-trascendental}
	M_a = E_a + \Sigma(E_a)  \mbox{ },
\end{equation}
is given by the sum of two terms: a bare energy, $E_a$, calculated within a potential model \cite{Godfrey:1985xj}, and a self energy correction, 
\begin{equation}
	\label{eqn:self-a}
	\Sigma(E_a) = \sum_{BC\ell J} \int_0^{\infty} k^2 dk \mbox{ } \frac{\left| M_{A \rightarrow BC}(k) \right|^2}{E_a - E_b - E_c}  \mbox{ },
\end{equation}
computed within the UQM formalism. 

Our results for the self energies of charmonia \cite{charmonium,Ferretti:2014xqa} and bottomonia \cite{bottomonium,Ferretti:2013vua,Ferretti:2014xqa} show that the pair-creation effects on the spectrum of heavy mesons are quite small. Specifically for charmonium and bottomonium states, they are of the order of $2 - 6\%$ and $1 \%$, respectively. 
The relative mass shifts, i.e. the difference between the self energies of two meson states, are in the order of a few tens of MeV. 
However, as QM's can predict the meson masses with relatively high precision in the heavy quark sector, even these corrections can become significant.
These results are particularly interesting in the case of states close to an open-flavor decay threshold, like the $X(3872)$ and $\chi_b(3P)$ mesons.
For example, in our picture the $X(3872)$ can be interpreted as a $c \bar c$ core [the $\chi_{c1}(2^3P_1)$], plus higher Fock components due to the coupling to the meson-meson continuum. In Ref. \cite{Ferretti:2014xqa}, we showed that the probability to find the $X(3872)$ in its core or continuum components is approximately $45\%$ and $55\%$, respectively.  

\section*{Acknowledgments}

This work is supported in part by PAPIIT-DGAPA, Mexico (grant IN107314).

\end{document}